\documentclass[twocolumn, 12pt]{iopart}

\usepackage{iopams}  
\usepackage{setspace}
\usepackage{multirow}\usepackage{pbox}
\usepackage{lineno,hyperref}
\modulolinenumbers[5]
\usepackage[dvipsnames]{xcolor}
\expandafter\let\csname equation*\endcsname\relax
\expandafter\let\csname endequation*\endcsname\relax

\usepackage{amsmath}
\usepackage{amssymb}
\usepackage{bm}
\usepackage{dcolumn}
\usepackage{float}
\usepackage{hyperref}
\usepackage{eso-pic}
\usepackage{ifthen}
\usepackage{atbegshi}
\usepackage{calc}

\usepackage{ tabularx }
\usepackage{ xcolor }
\usepackage{ cancel }
\usepackage{ rotating }
\usepackage{ siunitx }
\usepackage{ xspace }
\usepackage{ todonotes }

\usepackage{soul}

\newcommand{\MATLAB}{\textsc{Matlab\textsuperscript{\textregistered} }}

\newcommand{\move}{\textcolor{black}} 

\newcommand{\new}{\textcolor{black}}
\newcommand{\maybe}{\textcolor{black}}
\newcommand{\pink}{\textcolor{black}}
\newcommand{\pinkk}{\textcolor{black}}

\begin{document}

\title{Optimized signal deduction procedure for the MIEZE spectroscopy technique}

\author{J. K. Jochum}
\address{Heinz Maier-Leibnitz Zentrum (MLZ), Technische Universit\"at M\"unchen, D-85748 Garching, Germany}
\ead{jjochum@frm2.tum.de}

\author{L. Spitz}
\address{Heinz Maier-Leibnitz Zentrum (MLZ), Technische Universit\"at M\"unchen, D-85748 Garching, Germany}
\address{Physik Department, Technische Universit\"at M\"unchen, D-85748 Garching, Germany}

\author{C. Franz}
\address{Heinz Maier-Leibnitz Zentrum (MLZ), Technische Universit\"at M\"unchen, D-85748 Garching, Germany}
\address{J\"ulich Centre for Neutron Science JCNS-MLZ, Forschungszentrum J\"ulich GmbH Outstation at MLZ FRM-II, 85747 Garching, Germany}

\author{A. Wendl}
\address{Physik Department, Technische Universit\"at M\"unchen, D-85748 Garching, Germany}
\address{Heinz Maier-Leibnitz Zentrum (MLZ), Technische Universit\"at M\"unchen, D-85748 Garching, Germany}

\author{J. C. Leiner}
\address{Physik Department, Technische Universit\"at M\"unchen, D-85748 Garching, Germany}
\address{Heinz Maier-Leibnitz Zentrum (MLZ), Technische Universit\"at M\"unchen, D-85748 Garching, Germany}

\author{C. Pfleiderer}
\address{Physik Department, Technische Universit\"at M\"unchen, D-85748 Garching, Germany}
\address{Zentrum f\"ur QuantumEngineering (ZQE), Technische Universit\"at M\"unchen, D-85748 Garching, Germany}
\address{Munich Center for Quantum Science and Technology (MCQST), Technische Universit\"at M\"unchen, D-85748 Garching, Germany}

\author{O. Soltwedel}
\address{Physik Department, Technische Universit\"at M\"unchen, D-85748 Garching, Germany}
\address{Institut f\"ur Physik  Kondensierter Materie, Technische Universit\"at Darmstadt, D-64289 Darmstadt, Germany}



\date{\today}
\begin{abstract}
We report a method to determine the phase and amplitude of sinusoidally modulated event rates, binned into four bins per oscillation\new{, based on data generated at the resonant neutron spin-echo spectrometer RESEDA}.
The presented algorithm relies on a reconstruction of the unknown parameters. 
It omits a calculation intensive fitting procedure and avoids contrast reduction due to averaging effects. 
It allows the current data acquisition bottleneck \new{at RESEDA} to be relaxed by a factor of four \pink{and thus increases the potential time resolution of the detector by the same factor}. 
We explain the approach in detail and compare it to the established fitting procedures of time series having four and 16 time bins per oscillation. 
In addition we present the empirical estimates of the errors of the three methods and compare them to each other. 
We show that the reconstruction is unbiased, asymptotic, and efficient for estimating the phase. 
Reconstructing the contrast increases the error bars by roughly $10\%$ as compared to fitting 16 time binned oscillations. 
Finally, we give heuristic, analytical equations to estimate the error for phase and contrast as a function of their initial values and counting statistics.
\end{abstract}
 

\maketitle

\section{Introduction}

\begin{figure*}
    \includegraphics[width=0.9\linewidth]{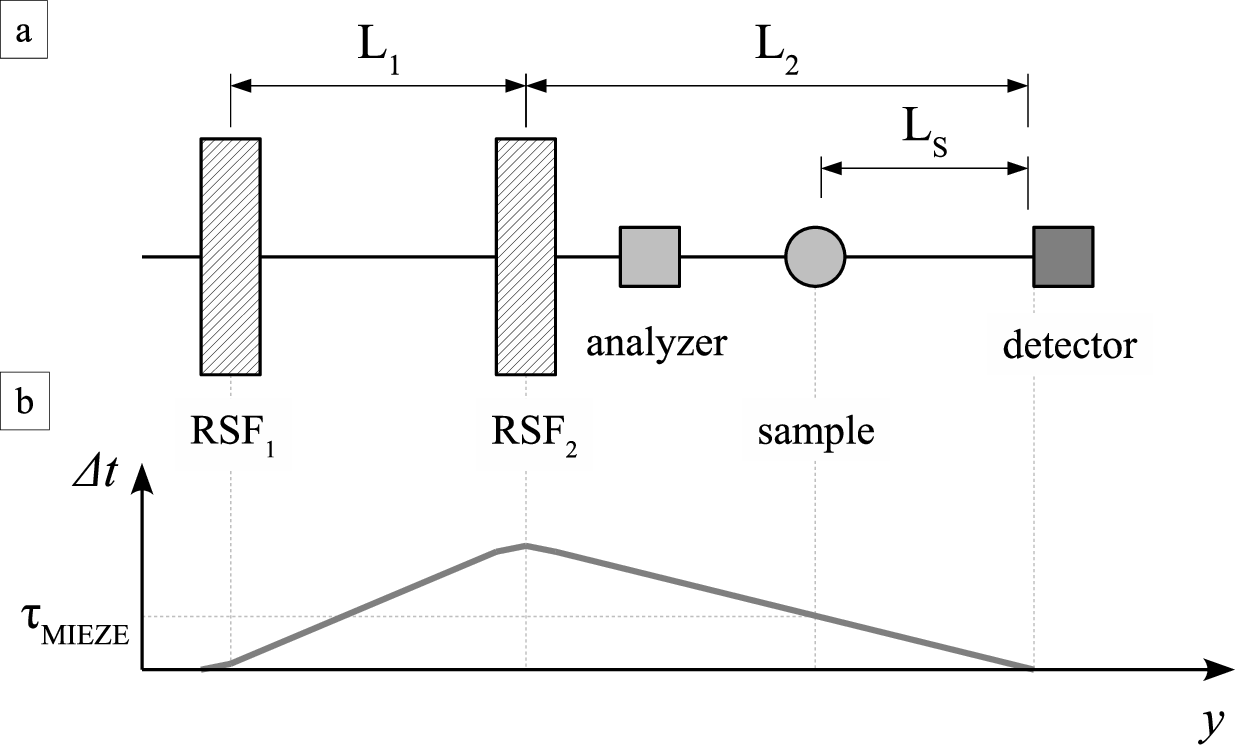}
    \caption{(a) Schematic representation of the essential parts of the MIEZE setup. Polarized neutrons travel in the y-direction passing the resonant spin flippers (RSF$_1$ and RSF$_2$) and the precession region between them, the spin analyser, the sample, and finally hitting the detector. (b) The time-of-flight difference $\Delta t$ of the spin eigenstates as function of distance along the flight path is shown. The time-of-flight difference at the sample position is $\tau_\mathrm{MIEZE}$}
    \label{pic:Scheme_MIEZE_4point}
\end{figure*}

MIEZE (Modulation of IntEnsity with Zero Effort) spectroscopy is a hybrid technique combining neutron resonance spin-echo and neutron Time-of-Flight spectroscopy.
It is routinely available at the spectrometer RESEDA at the Heinz Maier-Leibnitz Zentrum \cite{reseda} \pink{and BL06 at the J-PARC Materials and Life Science Experimental Facility \cite{2006Kawabata, 2013Hino}}. 
\pink{Furthermore, MIEZE is being actively developed at the Reactor Institute Delft, the ISIS neutron source \cite{2019Geerits} and Oak Ridge National Laboratories \cite{2020Dadisman}.}
In Fig. \ref{pic:Scheme_MIEZE_4point}\,a) we present a basic MIEZE setup.
It uses neutron spin precession generated by two resonant (neutron) spin flippers ($RSF_1$ and $RSF_2$) separated by a distance $L_\mathrm{1}$ and operated at individual frequencies ($f_1\,<\, f_2$), to manipulate the spin eigenstates \cite{1996Gahler}.
The resulting interference pattern of the superposition of the spin states corresponds to a sinusoidal intensity as a function of time akin to an optical heterodyne interferometer (see Fig. \ref{pic:Scheme_MIEZE_4point}\,b)).

The modulation frequency of this intensity is given by twice the difference of the $RSF$ frequencies $f_\mathrm{MIEZE}\,=\, 2(f_2-f_1)$ \cite{1998Felber, 2019Jochum}. In practice these frequencies are limited at the lower end by the neutron spin flip efficiency generated by the Bloch-Siegert shift to $f_\mathrm{min}\,=\,35\,$kHz \cite{1940Bloch}. 
The limitations at the upper end are due to skin and proximity effects in the resonant flippers, as well as parasitic capacities in the resonant circuits, which currently sets the maximum $RSF$ frequency to $f_\mathrm{max}\,=\,3.6\,$MHz \cite{2015Groitl, 2020Jochum}.

In contrast to conventional neutron spin-echo, the quantity measured in MIEZE corresponds to a sinusoidally modulated intensity in time, \move{from which the MIEZE contrast $C\,=\,\frac{I_0}{I_\mathrm{mean}}$, \new{with $I_\mathrm{mean}$ the time average intensity and $I_\mathrm{0}$ the amplitude of the intensity,} can be extracted }\cite{1992Gaehler}. 

\move{In order to increase the \pink{time} resolution (the Fourier time, $\tau_\mathrm{MIEZE}$), $f_\mathrm{MIEZE}$ has to be maximized since it is directly proportional to $\tau_\mathrm{MIEZE}$ via the following relationship:}
\begin{equation}
    \tau_\mathrm{MIEZE}\,= \frac{2\pi\,\hbar\,L_\mathrm{{S}}}{m_\mathrm{n} v_\mathrm{n}^3}\,f_\mathrm{MIEZE}
\end{equation}
\move{with the neutron mass $m_\mathrm{n}$, its velocity $v_n$, and the sample to detector distance $L_\mathrm{S}$ (cf. Fig. \ref{pic:Scheme_MIEZE_4point}).} \move{Further details and a description of the MIEZE setup may be found in Refs.} \cite{2019Franz, 2020Oda}.

\pinkk{The process of data reduction is done in two steps. In the first step, events are registered from electronic signals at the readout of the detector. These events are either accepted as neutron counts and then histrogrammed on the FPGA or rejected based on event length \cite{2010Schmidt}. 
In the second step, contrast and phase are deduced from this 4D histogrammed data set (pixel x pixel x time bins x foils). This is done by fitting the time bins, in each pixel and on every foil using a sine. 
Here, we present an approach that optimizes the second step of this procedure, requiring less computing power and allowing an increase in maximum achievable Fourier time by a factor of four.}

From a practical point of view the detector registers events per oscillation and histograms them according to a certain number of time bins \cite{2016Koehli}. 
Thus, for a fixed number of time bins, the length of each time bin is a function of the modulation frequency. 
The lower limit of the time bin length is given naturally by the temporal resolution of the detector, which is limited by the electron drift time and the clock of the electronics readout of the detector \cite{2016Koehli}. 
Hence, to detect signals with fast modulation, it is necessary for the number of time bins to be as low as possible. 
An insufficient number of time bins per oscillation however results in a smearing of the recorded oscillation amplitude, \new{and a loss in contrast. Therefore, an optimal compromise between the two needs to be achieved.}

Without loss of generality and neglecting the average background count rate, the event rate registered by a detector recording signals at discrete intervals in time is given by the integral over a harmonic oscillation with amplitude $I_0$ and arbitrary phase $\phi_0$:
\begin{subequations}
    \begin{align}
        I^{'}\,&=\,\frac{I_0}{\Delta\phi}\,\int_\mathrm{-\frac{\Delta\phi}{2}}^{\frac{\Delta\phi}{2}}\,\sin{(\phi-\phi_0)}\,d\phi    \label{eq1} \\
        I^{'}\,& =\,I_0 \sin(\phi_0)\,\frac{\sin{\frac{\Delta\phi}{2}}}{\frac{\Delta\phi}{2}}, \label{eq2}
    \end{align}
\end{subequations}
where $\Delta\phi\,=\,\frac{2\pi}{\#\,time bins}$. 
In this resolution function the sinc function acts as a damping factor, which assumes a minimum value if $\Delta\phi\,=\,0$, i.e. infinite time bins representing a trivial but trivially impractical solution.

\pink{Moreover, an infinite number of time bins would require infinite time-stamp accuracy of every event detected. Wrongly binned events decrease the contrast. This reduction scales with the ratio between time-stamp accuracy and time bin length. From this perspective fewer time bins are preferable as well.}

\pink{The final measurement quantity extracted from a MIEZE measurement is the intermediate scattering function $\mathcal{I}$($Q,\tau$) which is determined by dividing the sample contrast by an appropriate resolution contrast: $\mathcal{I}(Q,\tau) = \frac{C_{sample}}{C_{resolution}}$.}
Since all MIEZE measurements are normalized to the instrumental resolution function (which depends equally on the damping factor), the damping factor cancels out, and therefore does not need to be taken into account explicitly.
\new{Nevertheless, it is important to track the damping factor, to not increase the error bars of the contrast beyond a reasonable limit.}

Keeping in mind that at least three parameters $I_\mathrm{mean}$ (the time average), $I_\mathrm{0}$ (the amplitude), and $\phi_0$ (the arbitrary phase) must be extracted from the signal, a minimum of three time bins is necessary for an unambiguous reconstruction. 
\pink{In contrast to classical NSE it is not possible to use a $^3$He counter as a MIEZE detector. In fact, the detector requirements are quite demanding:
A MIEZE detector requires highest spatial and temporal resolution while in addition the thickness of the conversion volume of the detection system in the neutron flight direction must not exceed the size of the MIEZE group which decreases with increasing MIEZE time. \cite{2010Schmidt}}
Currently a \pink{CASCADE} detector with 16 time bins is used to detect the MIEZE signals at RESEDA \cite{2016Koehli}.
\pink{The detector consists of eight $^{10}$B coated detection foils, with a conversion layer thickness of 0.8-1.5\,$\mu$m and a pixel size of 1.56\,mm. 
The current CIPix ASIC preamplifier readout of the detector electronics is able to handle frequencies up to 10\,MHz. 
The recent improvements of the instrument RESEDA \cite{2020Jochum} have pushed the first generation CASCADE detectors to their limits. Nevertheless the time stamp accuracy of the detected events still have reserves, since the internal clock and the FPGAs run at a frequency of 40\,MHz, leading to a maximum binning inaccuracy of 12.5\,ns.}
\move{These constraints \pink{imposed by the detection system} limit the maximum MIEZE frequency at RESEDA to} $f_\mathrm{MIEZE}\,=\, \frac{10\,\mathrm{MHz}}{16}\,=\,625\,$kHz. 
In this regime, the damping induced by the sinc function is only 0.64\%.
\pink{The contrast is extracted from $I_\mathrm{mean}$, $I_\mathrm{0}$, and $\phi_0$, which are determined through a sine fit across the 16 time channels.} This fitting procedure is calculation intensive and cannot be performed in real time alongside the data acquisition.

To increase \pink{time resolution ($f_{\mathrm{MIEZE}}$)}, a practical solution is to apply the same routine with a reduced number of time bins.
Alternatively, one may find an unbiased estimate to reconstruct the parameters from the minimum necessary time bins by taking the time integration of the detector into account. 
In the following sections, a reconstruction procedure of the underlying parameters will be deduced using only four time bins.
This relaxes the required data collection interval by a factor of four corresponding to $f_\mathrm{MIEZE}^\mathrm{max}\,=\,2.5\,$MHz, thereby increasing the \pink{time} resolution by a factor of four.
Although three time bins are the optimal choice to cover the highest frequencies, we focus here on four time bins because of their backwards compatibility with older data sets histogrammed in 16 time bins. 

\section{Reconstruction of the MIEZE Signal}

As starting point for the reconstruction of the MIEZE signal, we give the mathematical description of the time dependent event rate $I(t)$ as recorded by the detector. 
This signal may be split into a time dependent and a time independent contribution ($I_\mathrm{mean}$). 
The latter describes the intrinsic background and all of the contrast reductions such as incoherent scattering, spin leakage, and sample dynamics. 
The sinusoidal time dependence is characterized by the amplitude $I_0$, the duration $T=\frac{1}{f_\mathrm{MIEZE}}$, and phase shift $\phi_0$. These combine to give $I(t)$ as:
\begin{equation}
    I(t)\,=\,I_\mathrm{mean}\,+\,I_0\,\sin\left(\frac{2\pi}{T}t-\phi_0\right).
\end{equation}

Since the time binning of events in the detector is equal to an integration over time of $I(t)$ in the respective interval, one may write the number of detected events in the $k^{th}$ interval $I_\mathrm{k}$ as:
\begin{equation}
    I_k\,=\,\frac{1}{T}\int_\mathrm{\frac{T}{N}(k-1)}^{\frac{T}{N}k}I(t) dt \label{eq3},
\end{equation}
 with $k = 1,2,3, ..., N$ for $N$ bins.
 Normalizing $I_\mathrm{k}$ by $I_\mathrm{mean}$ corresponds to the probability of a single event occurring in the $k^{th}$ interval.

\begin{figure*}[ht!]
\includegraphics[width=0.9\linewidth]{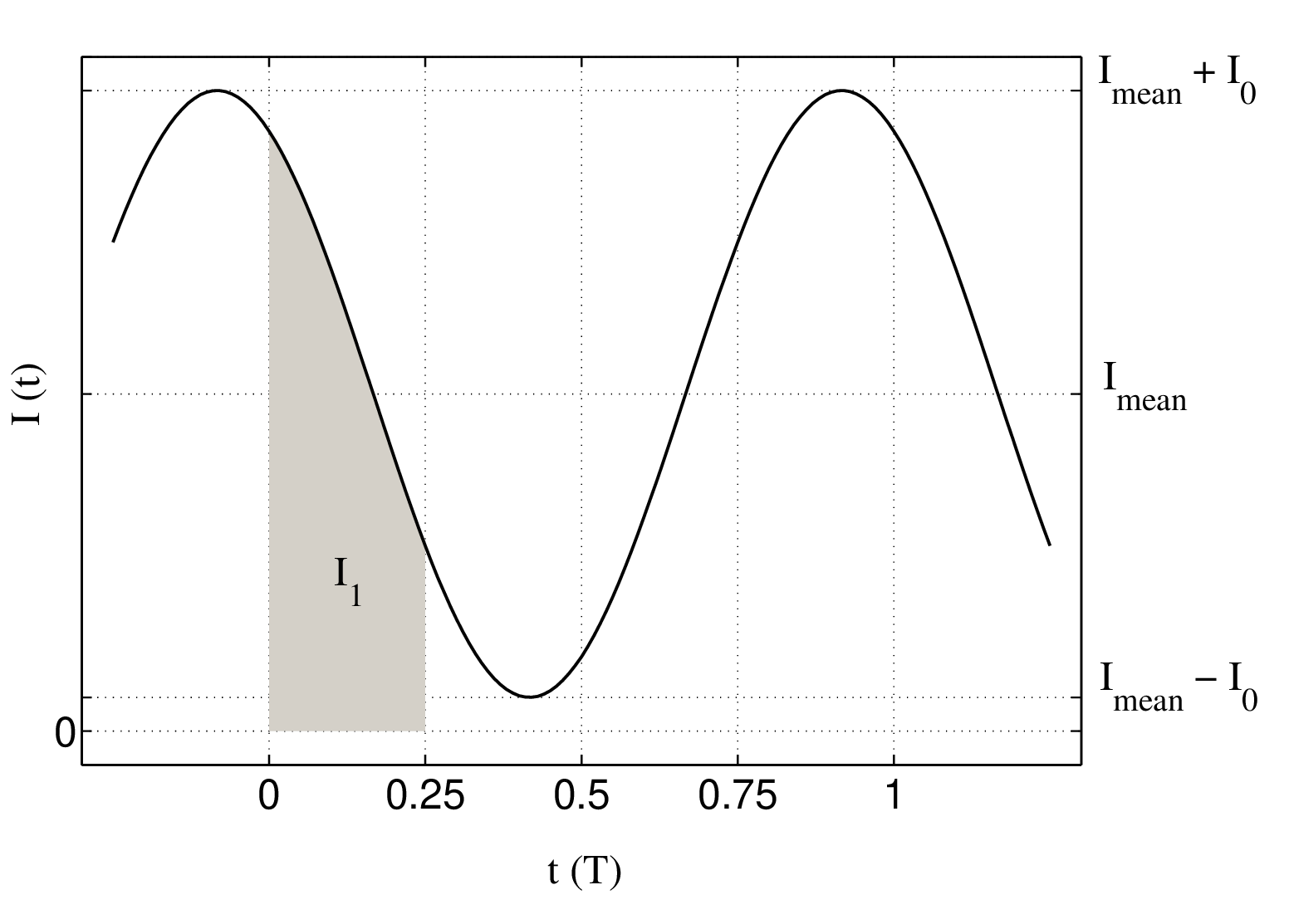}
\caption{ A typical time dependent sinusoidal intensity variation with phase $\phi_0\,=\,\frac {\pi}{8}$ and a contrast $C\,=\,\frac{I_0}{I_\mathrm{mean}}$ that is defined by the amplitude $I_0$ and the mean value $I_\mathrm{mean}$. The gray shaded area $I_1$ normalized to $I_\mathrm{mean}$ is the probability of a single event being detected in the first interval from the division of each oscillation of $I(t)$ into four equally long time bins.}
\label{pic:Skizze}
\end{figure*}

For a subdivision into four intervals ($N\,=\,4$ cf. Fig.\,\ref{pic:Skizze} gray shaded area for $I_1$) one may rewrite \eqref{eq3} as follows:

\begin{subequations}
    \begin{align}
        I_1\,&=\,\frac{I_\mathrm{mean}}{4}+\frac{I_0}{2\pi}\left( \sin(-\phi_0)+\cos(-\phi_0)\right), \label{subeq1}\\
        I_2\,&=\,\frac{I_\mathrm{mean}}{4}+\frac{I_0}{2\pi}\left( \cos(-\phi_0)-\sin(-\phi_0)\right), \label{subeq2}\\
        I_3\,&=\,\frac{I_\mathrm{mean}}{4}+\frac{I_0}{2\pi}\left(-\sin(-\phi_0)-\cos(-\phi_0)\right), \label{subeq3}\\
        I_4\,&=\,\frac{I_\mathrm{mean}}{4}+\frac{I_0}{2\pi}\left(-\cos(-\phi_0)+\sin(-\phi_0)\right) \label{subeq4}.
    \end{align}
\end{subequations}

Summing up neighbouring intervals and simplifying the results yields:

\begin{subequations}
    \begin{align}
        I_1+I_2\,&=\,\frac{I_\mathrm{mean}}{2}+\frac{I_0}{\pi} \cos(-\phi_0), \label{subeq5}\\
        I_2+I_3\,&=\,\frac{I_\mathrm{mean}}{2}-\frac{I_0}{\pi} \sin(-\phi_0), \label{subeq6}\\
        I_3+I_4\,&=\,\frac{I_\mathrm{mean}}{2}-\frac{I_0}{\pi} \cos(-\phi_0), \label{subeq7}\\
        I_4+I_1\,&=\,\frac{I_\mathrm{mean}}{2}+\frac{I_0}{\pi} \sin(-\phi_0) \label{subeq8}.
    \end{align}
\end{subequations}

Adding the next nearest neighbour intervals (I$_{k}$ + I$_{k+2}$) yields only the direct component (first terms) while the phase information is lost:
\begin{equation}
    I_{k}+I_{k+2}\,=\,\frac{I_\mathrm{mean}}{2}.
\end{equation}
This is a direct consequence of the signal's harmonic periodicity.\\

Since equations\,\eqref{subeq5}\,-\,\eqref{subeq8} are sums of neighbouring intervals, \new{one may use two independent but identical detector read outs to measure two separate time intervals that are $\frac{\pi}{2}$ phase shifted relative to each other. This yields equivalent information, but allows for a doubling of $f_\mathrm{MIEZE}^\mathrm{max}$.} \\

\pink{Precise phase determination of harmonic signals is} well established \pink{using} quadrature detection \pink{in} optical interferometry \cite{2012Rerucha} \pink{or signal processing} \pink{where $\pi/2$ phase shifted signals (}\,\eqref{subeq5} - \eqref{subeq8}\pink{) are} combined to \move{reconstruct the \pink{unknown} phase} $\phi_\mathrm{0}$:

\begin{equation}
    \tan(-\phi_\mathrm{0})\,=\,\frac{I_\mathrm{4} + I_\mathrm{1} - (I_\mathrm{2} + I_\mathrm{3} )} {I_\mathrm{1} + I_\mathrm{2} - (I_\mathrm{3} + I_\mathrm{4} )} \label{phase}.
\end{equation}

It is also possible to deduce the phase by subtracting equations\, \eqref{subeq1}\,-\,\eqref{subeq4} from each other: 

\begin{equation}
        \frac{I_\mathrm{1} - I_\mathrm{2}}{I_\mathrm{1} - I_\mathrm{4}}\,=\,\frac{I_\mathrm{1} - I_\mathrm{2}}{I_\mathrm{2} - I_\mathrm{3}}\,=\,\frac{I_\mathrm{4} - I_\mathrm{3}}{I_\mathrm{1} - I_\mathrm{4}}\,=\,\frac{I_\mathrm{4} - I_\mathrm{3}}{I_\mathrm{2} - I_\mathrm{3}}\,=\,\tan(-\phi_\mathrm{0}). \label{subeq12}
\end{equation}

Eq.\,\ref{subeq12} shows that in principle one interval can be neglected.
However, for this approach information in the form of counts is ignored within that interval, thus reducing the overall statistics and accuracy. 
The average over equation\,\eqref{subeq12} equals equation\,\eqref{phase}:

\new{Using $C = \frac{I_0}{I_\mathrm{mean}}$ the reconstructed (rec) contrast may be deduced as well,} by combining either equations\,\eqref{subeq5} and \eqref{subeq7} or \eqref{subeq6} and \eqref{subeq8}:

\begin{subequations}
    \begin{align}
        C_\mathrm{1, rec}\,=\,\frac{I_1 + I_2 - (I_3 + I_4)} {I_1 + I_2 + I_3 + I_4} \cdot \frac{\pi}{2 \cos{\phi_\mathrm{0}}}, \label{subeq13} \\
        C_\mathrm{2, rec}\,=\,\frac{I_1 + I_4 - (I_2 + I_3)} {I_1 + I_2 + I_3 + I_4} \cdot \frac{\pi}{2 \sin{\phi_\mathrm{0}}}. \label{subeq14}        
    \end{align}
\end{subequations}

Of course the accuracy of the evaluated contrast is strongly coupled to the accuracy of the estimated phase and diverges at the singularities, i.e., when $\cos{\phi_\mathrm{0}}$ or $\sin{\phi_\mathrm{0}}$ tend towards zero. 
In order to avoid the singularities we apply equations\,\eqref{subeq13} and \eqref{subeq14} for the appropriate case:

\begin{equation}
C_\mathrm{rec} =
  \begin{cases}
    C_\mathrm{1,rec},       & \quad \text{for   } \cos{\phi_\mathrm{0}} \geq \sin{\phi_\mathrm{0}}\\
    C_\mathrm{2,rec},       & \quad \text{for   } \cos{\phi_\mathrm{0}} < \sin{\phi_\mathrm{0}}\label{subeq16a}
  \end{cases}
  .
\end{equation}

\maybe{We emphasize again that this simple reconstruction of contrast and phase using only four time bins, allows an increase in Fourier time by a factor of four. Additionally, this reconstruction method (unlike the previously used method) does not require any computationally intensive fitting, which will speed up data reduction immensely and allow for real time data reduction, which will make it possible to optimize measuring times, and use allocated beamtime more efficiently.}

\section{Estimation of the confidence interval}

Next, we discuss how many events are necessary to determine phase and contrast with a desired accuracy.
\new{For this we will compare three different attempts: 1) The 16 time bin fitting method used so far at RESEDA \textcolor{blue}{(fit,16)}, 2) the four time bin fitting method \textcolor{ForestGreen}{(fit,4)}, 3) the four time bin reconstruction method \textcolor{red}{(rec)}.}
\pink{The procedure does not take into account a possible phase jitter of the detector signal.}
\new{The \MATLAB code utilized for these calculations has been made available for reference \cite{MatlabCode}.}
As a first attempt, the uncertainties are estimated using Gaussian error propagation with the relative errors ${\frac{1}{\sqrt{I_k}}}$. 
Deducing the partial derivatives is straightforward. Less obvious is the estimate of the total errors $\Delta I_1,...\,I_4$, due to their mutual dependence. Moreover, the total errors also depend on $C_0$ and $\phi_0$.
\new{It needs to be mentioned that, while a generalized Gaussian error propagation would account for the covariance between all parameters, it is not able to give a reliable answer in the limit I$\rightarrow$0. Therefore,}
we applied simulations and executed them for various initial phases ($\phi_\mathrm{0} =\ang{0}, \ang{15}, ..., \ang{120}$) and contrasts ($C_\mathrm{0} = 0.05, 0.1, ..., 0.95$).
First 10 single events with the desired sinusoidal distribution are generated using the pseudo-random generator of \MATLAB and histogrammed subsequently. 
For a given $C_0$ and $\phi_0$, the probability to fall in a certain time bin is determined by equations\,\eqref{subeq1}\,-\,{\eqref{subeq4}}. 
Subsequently, the phase and contrast are calculated according to the \new{three different methods}. 
Next, new events are added to this \pink{run} and the evaluation is repeated recursively. 

\begin{figure*}[ht!]
\includegraphics[width=1.05\linewidth]{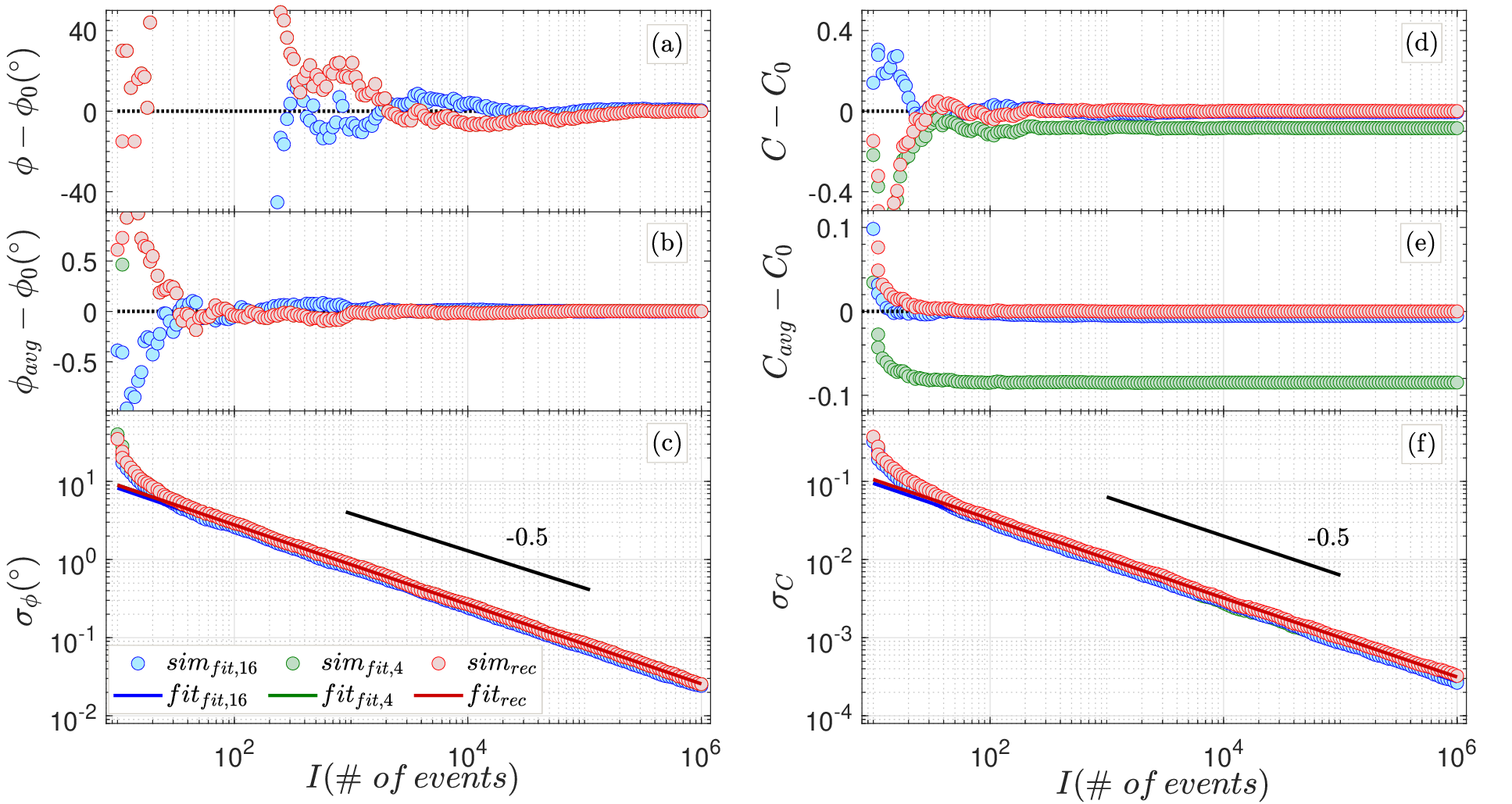}
\caption{For the initial parameters $C_\mathrm{0} = 0.85$, $\phi_\mathrm{0} = \ang{60}$, approximated deviations for phase $\phi$ ($(a)$,$(b)$,$(c)$) and contrast $C$ ($(d)$,$(e)$,$(f)$) versus the number of total count events ($I\,=\,I_1 + I_2 + ... + I_k$) for a single \pink{run} ($(a)$ and $(d)$) and averaged over 500 \pink{runs} ($(b)$ and $(e)$). The standard deviations ($(c)$ and $(f)$) calculated respectively for the reconstruction method \textcolor{red}{(rec)} and generic fitting procedures using four \textcolor{ForestGreen}{(fit,4)} and 16 \textcolor{blue}{(fit,16)} time bins are displayed as dots together with their fits (solid lines) \pink{from which the power law exponents were extracted. For clarification} the black lines show a power law with an exponent of $-0.5$. To highlight the significance of the resulting phase as a function of events, $(a)$ and $(b)$ have been purposefully cropped. Except for Fig.\,\ref{pic:Simu_Results_Phase_Contrast_c0p85_60deg}  $(d)$ and $(e)$, the green data points coincide with the red data points reflecting nearly identical values.}
\label{pic:Simu_Results_Phase_Contrast_c0p85_60deg}
\end{figure*}

The number of added events in such a series increases logarithmically.
This ensures a low computational burden over a large dynamic range of events (here over five orders of magnitude) and keeps the evaluation equally weighted in a logarithmic representation.
Finally, the results are compared with each other. 
\pink{Figures \ref{pic:Simu_Results_Phase_Contrast_c0p85_60deg} (a) and (d) show the phase ($\phi$) and contrast ($C$) for one run.}
It is found that the phases estimated for the four point fitting method (green) and the reconstruction (red) are identical within error for more than 30 events.

For a low number ($< 30$) of events, the phase and contrast values have larger deviations from the true values as the result of insufficient statistics.
As expected from equation\,\eqref{eq2}, both fitting methods show biased (damped) contrast estimates. 
For the contrast $C_0\,=\,0.85$ presented in Fig. \ref{pic:Simu_Results_Phase_Contrast_c0p85_60deg} \pink{(d) and \ref{pic:Simu_Results_Phase_Contrast_c0p85_60deg} (e)}, the expected damping according to Eq.\,\eqref{eq2} is $0.64\%\cdot C_0\,=\,0.0054$ for the 16 time bins (blue) and $10\%\cdot C_0\,=\,0.085 $ for four time bins (green).
This shows that the estimates inferred from the reconstruction method are unbiased.

To estimate the standard deviations of the phase and the contrast, the simulation was \pink{run} 500 times.

From these data, the average phase ($\phi_\mathrm{avg}$) and contrast ($C_\mathrm{avg}$) as well as their corresponding standard deviations ($\sigma_\mathrm{\phi}$ and $\sigma_\mathrm{C}$) were calculated as a function of events $I$ (cf. Fig.\,\ref{pic:Simu_Results_Phase_Contrast_c0p85_60deg} (b), (c), (e) and (f)).
While the average phase is estimated correctly, the unbiased estimate for the contrast bears the expected damping. \move{In agreement  with the experimental behaviour,} the estimated standard deviations $\sigma_\mathrm{\phi}$ and $\sigma_\mathrm{C}$ (for the reconstruction and fitting procedures) decrease with the same asymptotic behavior as the total number of events ($I\,=\,\sum_\mathrm{k}^{N}{I_k}$) increases (cf. Fig.\,\ref{pic:Simu_Results_Phase_Contrast_c0p85_60deg} (c) and (f)). This proves that the applied estimator is consistent.

The relationship between standard deviation and events, for both the phase and contrast, is described by simple power laws:
\begin{subequations}
    \begin{align}
        \sigma_\mathrm{\phi}(I)\,&=\,10^{\beta_\mathrm{\phi}}\cdot I^{\alpha_\mathrm{\phi}},   \label{subeq15} \\
        \sigma_\mathrm{C}(I)\,&=\,10^{\beta_\mathrm{C}}\cdot I^{\alpha_\mathrm{C}}.            \label{subeq16}
    \end{align}
\end{subequations}

From a linear fit to the log-log plot of the estimated standard deviations (c.f. Fig.\,\ref{pic:Simu_Results_Phase_Contrast_c0p85_60deg} panels (c) and (f)) for more than 30 events, the power law exponents ($\alpha_\mathrm{\phi}$ and $\alpha_\mathrm{C}$) may be inferred:

\begin{equation}
    \alpha_\mathrm{\phi}\,=\,\alpha_\mathrm{C}\,\thickapprox\,-0.5 . \label{eq4}
\end{equation}


\begin{figure*}[ht!]
\includegraphics[width=1.05\linewidth]{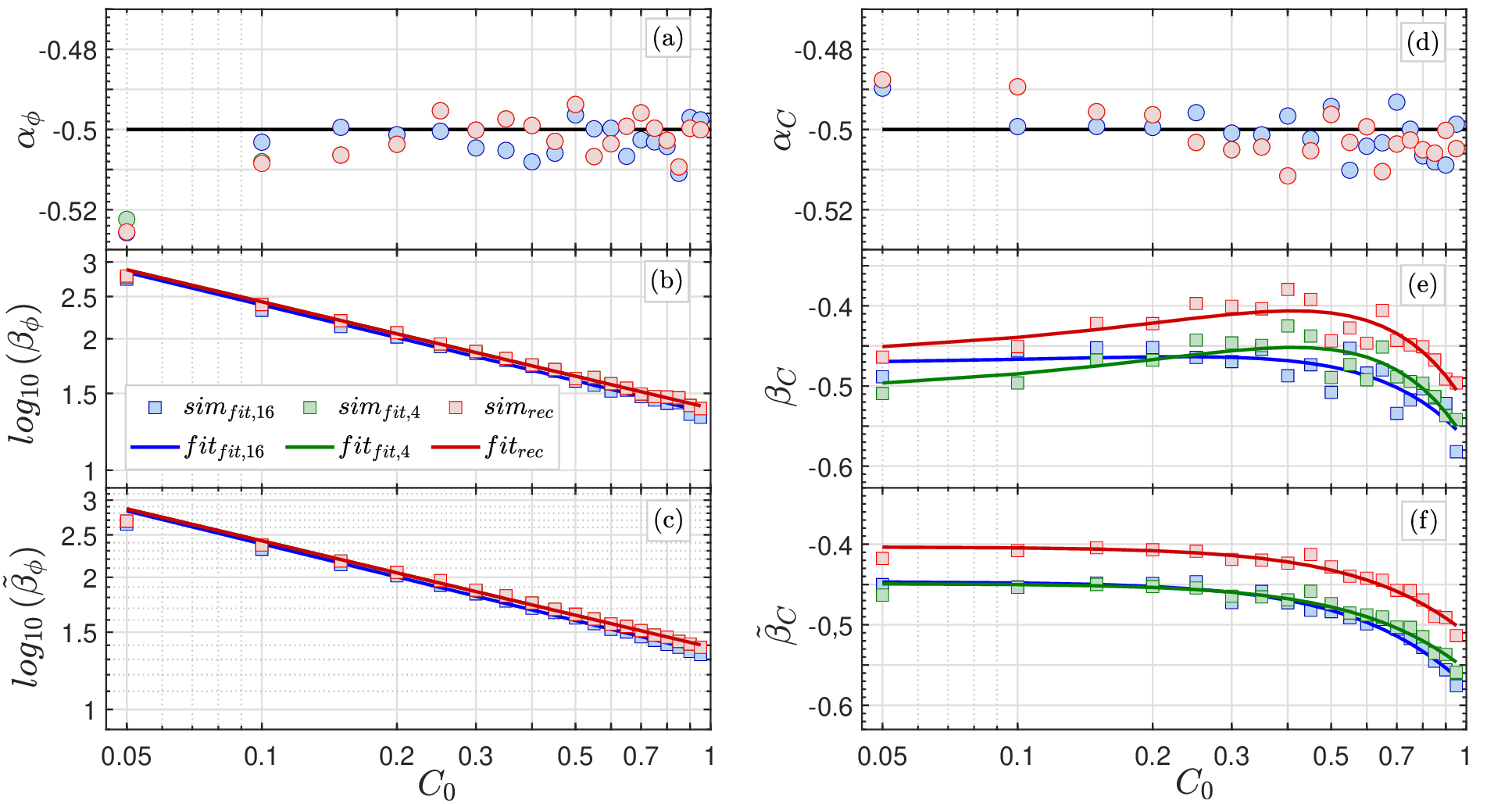}
\caption{Fit parameter of the error estimation for phase ($\alpha_\mathrm{\phi}$ (a) and $\beta_\mathrm{\phi}$ (b)) and contrast ($\alpha_\mathrm{C}$ (d) and $\beta_\mathrm{C}$ (e)) as a function of the initial contrast $C_\mathrm{0}$ and a fixed phase $\phi_\mathrm{0}\,=\,\ang{60}$. Panels (c) and (f) show the scaling parameters $\tilde{\beta}_{\phi}$ and $\tilde{\beta}_{C}$ using the constraint $\alpha_{*}\,=\,-0.5$. The color code is the same as for Fig.\,\ref{pic:Simu_Results_Phase_Contrast_c0p85_60deg}: \pink{reconstruction method \textcolor{red}{(rec)}, generic fitting procedures using four \textcolor{ForestGreen}{(fit,4)} and 16 \textcolor{blue}{(fit,16)} time bins}. Apart from plots $(e)$ and $(f)$, the values deduced from four-point-fit and reconstruction method overlap \pink{with} each other.}
\label{pic:p_Results_60deg}
\end{figure*}

To test the generality of this power law behavior and to determine the missing parameters ($\beta_\mathrm{\phi}$ and $\beta_\mathrm{C}$) for varying contrasts, the simulations for $C_\mathrm{0}\,=\,0.05, 0.1, ..., 0.95$ were repeated while keeping the initial phase fixed $\phi_\mathrm{0}\,=\,\ang{60}$.
In Fig.\,\ref{pic:p_Results_60deg}, parameters of the reconstruction and fitting methods are deduced for a comprehensive range of representative contrasts.
For each such contrast ($C_\mathrm{0}$), ${\alpha_\mathrm{\phi}}$ and ${\alpha_\mathrm{C}}$ (cf. \ref{pic:p_Results_60deg} (a) and (d)) remain nearly unchanged, confirming that the use of a normal distribution to approximate a Poisson distribution is well justified.
\move{However}, the coefficients $\beta_\mathrm{\phi}$ and $\beta_\mathrm{C}$ show quantitatively distinct dependencies on the initial parameter $C_\mathrm{0}$ (cf. Fig. \ref{pic:p_Results_60deg} (b) and (e)). 
\new{$\beta_{\phi}$} was observed to follow an exponential decay with increasing $C_0$:
\begin{equation}
    \beta_{\phi}\,=\,\beta_{2,\phi} C_{0}^{\beta_{1,\phi}},     \label{eq5}
\end{equation}
with decay constants $\beta_{2,\phi}$ and $\beta_{1,\phi}$ which vary slightly depending on the method. 
The functional dependence of the contrast is less obvious, and the parabolic fits (solid lines) in Fig.\,\ref{pic:p_Results_60deg} (e) are not ideal.

\begin{figure*}[ht!]
\includegraphics[width=1.05\linewidth]{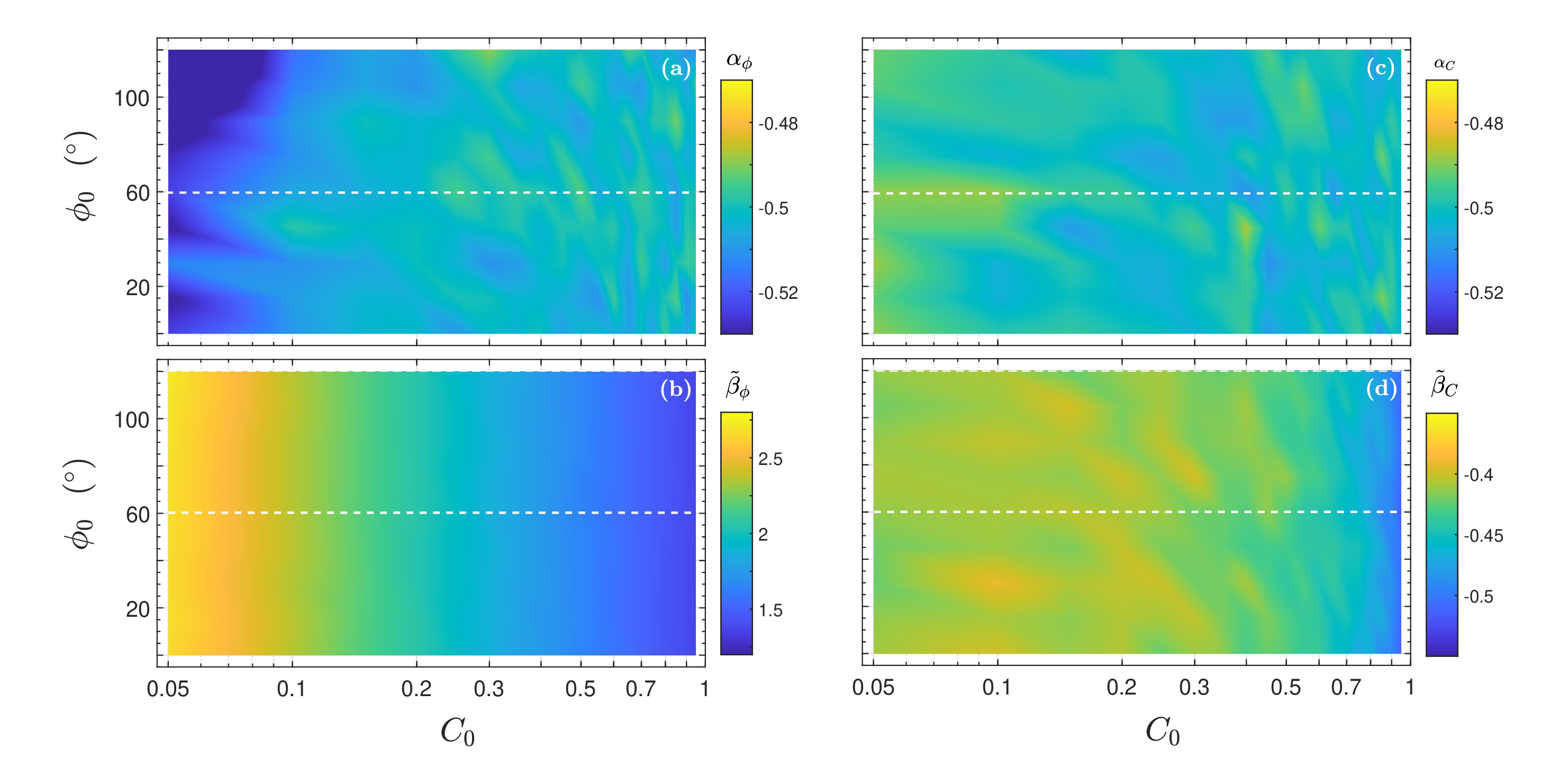}
\caption{\new{Color-map plots of parameters $\alpha_\mathrm{\phi}$ (a), $\tilde{\beta_\mathrm{\phi}}$ (b), $\alpha_\mathrm{C}$ (c), $\tilde{\beta_\mathrm{C}}$ (d) for varying initial phases $\phi_0$ and contrasts $C_0$ using the reconstruction. All four parameters are nearly independent of the initial phase $\phi_\mathrm{0}$. While the parameters ($\alpha_\mathrm{\phi}$ and $\alpha_\mathrm{C}$) remain nearly constant even through varying initial contrasts, $\tilde{\beta_\mathrm{\phi}}$ decays exponentially and $\tilde{\beta_\mathrm{C}}$ shows the same trend as shown in Fig. \ref{pic:p_Results_60deg} (c) for increasing $C_0$}} \label{pic:ParameterPlot}
\end{figure*}

Since $\alpha_{*}$ and $\beta_{*}$ ($*\,=\,\phi$ or $C$) depend on each other, as the fits are over-parameterized, $\beta_{*}$ was recalculated with the constraint $\alpha_{*} = -0.5$.
For the sake of clarity, $\beta_{*}$ is renamed $\tilde{\beta}_{*}$ in the following if the constraint $(\alpha_{*}\,=\,-0.5)$ is applied.
The resulting fits are plotted in Figs. \ref{pic:p_Results_60deg} (c) and (f). 
Compared to Fig. \ref{pic:p_Results_60deg}\,(b), The exponential dependence of $\tilde{\beta}_{\phi}$ is maintained.
Furthermore, $\tilde{\beta}_{C}$ can now be described well by a shifted half-normal distribution:

\begin{equation}
 \tilde{\beta}_{C}\,=\,\beta_{2,C} \cdot e^{-\left(\frac{C_0}{\beta_{1,C}}\right)^2}-1 .\label{eq6}
\end{equation}


To show that these findings hold for the relevant range of phases, this procedure was repeated for $\phi_0\,=\,(0...\,\ang{120})$ in steps of $\ang{15}$. 
\new{To confirm the $\ang{90}$ periodicity of the angular dependence, the interval was extended to $\ang{120}$.}
This yields a set of curves comparable to the ones in Fig. \ref{pic:p_Results_60deg}, which are color-plotted in Fig. \ref{pic:ParameterPlot}, highlighting their behavior throughout the entire parameter space.
The plots confirm that the fitting parameters deduced with these techniques are practically independent of $\phi_0$. 
One may note that due to the periodicity of the harmonic functions, these findings are valid for all phases.

Combining equations\,\ref{subeq15}, \ref{subeq16}, \ref{eq5}, and \ref{eq6}, we find the analytical equations for the estimate of the standard deviation:

\begin{subequations}
    \begin{align}
        \sigma_\mathrm{\phi}\,&=\,10^{\left(\beta_\mathrm{2,\phi}\cdot C_0^{\beta_\mathrm{1,\phi}}\right)}\,\cdot\,\frac{1}{\sqrt{I}},  \label{subeq17} \\
        \sigma_\mathrm{C}\,&=\,10^{\left(\,\beta_\mathrm{2,C}\,\cdot\,e^{-\left(\frac{C_0}{\beta_{1,C}}\right)^2}\,-\,1\right)} \cdot\, \frac{1}{\sqrt{I}}. \label{subeq18}
    \end{align}
\end{subequations}

\new{The parameters $\beta_{1,\phi}$, $\beta_{2,\phi}$, $\beta_{1,C}$, and $\beta_{2,C}$ we found for the different methods presented here} are summarized in Table \ref{tab:parameter}.

\begin{table*}[ht!]
\centering
\caption{Parameters to deduce the standard deviations $\sigma_{\phi}$ and $\sigma_C$ using equations\,\ref{subeq17} and \ref{subeq18}, for the reconstruction and four- and 16-point fitting methods.} 
\label{tab:parameter}
\begin{tabular}{||c||c|c||c|c||} \hline 
$method$ &  $\beta_{1,\phi}$ & $\beta_{2,\phi}$ & $\beta_{1,C}$ & $\beta_{2,C}$\\ \hline \hline
reconstruction & -0.244 & 1.383 & 2.29 & 0.60 \\ \hline
four-point-fit & -0.244 & 1.341 &  7.90 & 0.95\\ \hline
16-point-fit & -0.250 & 1.383 & 9.15 & 0.95\\ \hline
\end{tabular}
\end{table*}

\new{Equations \ref{subeq17} and \ref{subeq18} and Table \ref{tab:parameter} show that the deduced errors depend strongly on the initial contrast $C_0$ and the applied methods. The most obvious variation is observed for the parameters $\beta_{1,C}$ and $\beta_{2,C}$. $\beta_{1,C}$ determines how quickly $\sigma_C$ drops with increasing initial contrast whereas $\beta_{2,C}$ scales the absolute magnitude of $\sigma_C$.}
\move{We would like to emphasize that the error bars deduced for the contrast using the four point fitting method must be treated carefully, since the procedure of inferring the estimate is biased. Re-scaling this contrast and its error with the damping factor of $0.9$ given by \eqref{eq1}, the same error observed for the reconstruction method is maintained.}
\pink{However, as long as the same procedure is used for sample and resolution measurements, these effects cancel out and can therefore be neglected.}

\section{Conclusions}

We have presented an algorithm to deduce the contrast and the phase of a sinusoidally modulated time series sampled at four data points per oscillation. 
Both contrast and phase are recovered in agreement with 16 time bins.  
\pink{The methods presented here are adequate to estimate phase and contrast of MIEZE signals. Intrinsically all three methods get less accurate in determining the phase as the contrast decreases. On the other hand, their accuracy is independent of the initial phase.}
The reconstruction trades in a higher time resolution for less accurate contrast. 
Quantitatively, this factor is better than $\frac {\sigma_\mathrm{C0, fit,16}}{\sigma_\mathrm{C0, rec}}\,\geq\,0.9$ compared to the fitting method, but may be compensated by increased statistics, i.e around 20\% prolonged counting time.
However, using the reconstruction, there is no fitting procedure involved which significantly reduces the required computational burden.
Thus, this method may be readily applied to a large number of detector pixels as the measurements proceed in time.
\new{As mentioned above, real time data evaluation will lead to a more efficient use of measurement time, \pink{decreasing the time needed for each experiment}.}

Most importantly, this new method solves one of the main limitations afflicting the MIEZE resolution. 
Using a CASCADE-type detector \cite{2016Koehli} with a maximum time resolution of 100\,ns (10\,MHz), the maximum intensity modulation frequency for 16 time channels is 625\,kHz, which, at 6\,\AA\, with the current dimensions at RESEDA yields a MIEZE (Fourier) time of $\sim$\,3\,ns.
In stark juxtaposition, the resolution limit using the four-point method is $\sim$12\,ns at 6\,\AA\, or $\sim$\,100\,ns at 12\,\AA\,.

\new{Having extended the \pink{time} resolution limit using the four point reconstruction method, the next challenge for MIEZE data acquisition is the pixel size of the detector.}
\move{The reason for this is that the coherence volume of the MIEZE signal is indirectly proportional to the wavelength of the incoming neutron beam, the width of the wavelength band, and most importantly $f_\mathrm{MIEZE}$.} 
Thus, for intensity modulation frequencies at or above 2.5\,MHz, extremely flat detector surfaces are needed to minimize phase differences within a single pixel. 
A $^{10}$B layer on a solid surface instead of Kapton foil could be a possible solution. Furthermore, a spherical detector foil shape would suppress the phase rings which occur on flat surfaces due to variations of path lengths \cite{2019Schober}.

\section{Acknowledgements}

We wish to thank B. Pompe from University of Greifswald and T. Keller from Max-Planck-Institute for Solid State Research Stuttgart for very useful discussions and M. Klein from CDT GmbH for discussions and support of the CASCADE detector system. Financial support through the BMBF projects ‘Longitudinale Resonante Neutronen Spin-Echo Spektroskopie mit Extremer Energie-Aufl\"{o}sung’ (F\"{o}rderkennzeichen 05K16W06) and 'Resonante Longitudinale MIASANS Spin-Echo Spektroskopie an RESEDA' (F\"{o}rderkennzeichen 05K19W05) is gratefully acknowledged.

\section{References}

\bibliographystyle{elsarticle-num}
\bibliography{ref.bib}

\end{document}